\documentclass{aa}

\usepackage{color}

\usepackage{graphicx}
\usepackage{txfonts}
\usepackage{natbib}

\bibpunct{(}{)}{;}{a}{}{,} 
\begin{document}


\title{Local stability and global instability in iron-opaque disks}
\author{Miko\l{}aj Grz{\c e}dzielski (mikolaj@cft.edu.pl) \inst{1}
\and
Agnieszka Janiuk
\inst{1}
\and
Bo{\.z}ena Czerny
\inst{1}}
\institute{Center for Theoretical Physics, Polish Academy of Sciences, Al. Lotnikow 32/46, 02-668 Warsaw, Poland}
\date{Received ...; accepted ...}
\abstract{The thermal stability of accretion disk and the possibility to see a limit-cycle behaviour strongly depends on the ability of the disk plasma to cool down. Various processes connected with radiation-matter interaction
appearing in hot accretion disk plasma contribute to opacity. For the case of geometrically thin and optically thick accretion disk, we can
estimate the influence of several different components of function $\kappa$, given by the Roseland mean. In the case of high temperatures ($\sim 10^7$) K, the electron Thomson
scattering is dominant. At lower temperatures atomic processes become important. The slope $d \log \kappa / d \log T$ can have locally stabilizing or destabilizing effect on the disk. Although the local MHD simulation postulate the stabilizing
influence of the atomic processes, only the global time-dependent model can reveal the global disk stability range estimation. This is due to global diffusive nature of that processes.
In this paper, using previously tested GLADIS code with modified prescription of the viscous dissipation, we examine the stabilizing effect of the \textit{Iron Opacity Bump}.
}
\authorrunning{Grzedzielski et al.}
\titlerunning{Local stability}
\maketitle
\section{Introduction}
The energy output observed in the Galactic X-ray Binaries (XRB), and Active Galactic Nuclei (AGN), suggests that the source of emitted power in these sources must be connected with the gravitational potential energy of a compact object.
In most of the former, and all of the latter, this object is a black hole. Its mass can range from a few solar masses up to a few billion of $M_{\odot}$.
The material that is accreted onto a black hole and emits radiation, may posses substantial angular momentum. In this case, the accretion flow forms a geometrically thin
disk, which is located in the equatorial plane of an XRB, or co-aligned with the plane perpendicular to the black hole rotation axis in AGN, if the latter is powered by a Kerr black hole.
The classical solution of \citet{SS1973} with their $\alpha-$viscosity prescription, describes a stationary accretion disk where the dissipated heat is
balanced by thermal radiation. As studied by \citet{PRP1973,LE1974} and \citet{SS1976}, the viscous stress tensor scaling with a total (i.e., gas plus radiation) pressure leads to the runaway instability of the disk structure. Alternatively, the viscosity prescription may be given by
a gas pressure only (in this case the thermal instability does not develop), or by some intermediate law, that invokes a combination of gas and radiation
with different weights (see e.g. \citet{Szuszkiewicz1990}). For instance, a general prescription that was recently discussed by \citet{viscositypaper} reads:
\begin{equation}
\tau_{r \phi} = \alpha P_{\rm tot}^{\mu} P_{\rm gas}^{1-\mu} ~.
\end{equation}
The above model leads to the unstable disk behaviour, which manifests in a limit-cycle type of oscillations of the emitted luminosity, characteristic for the periodically heated and cooled inner regions of the accretion disk. This kind of cycle is possible, if only the thermal runaway is captured by some stabilizing process. This might be advection of heat
onto a black hole, as proposed for the so-called 'slim disk' solution \citep{Abramowicz1988}.
The presence of radiation pressure instability in action of cosmic sources has been a matter of debate (see e.g.,  review by \citet{Blaes2014}). Nevertheless, there
are strong observational hints which support the limit-cycle type of behaviour in at least two well-known microquasars, GRS 1915+105 and IGR J17091-324, in some of their spectral states \citep{Belloni2000,Altamirano2011}.
The limit-cycle oscillations were detected also in the Ultraluminous X-ray source HLX-1, claimed to contain an intermediate-mass black hole \citep{Farrell2009,Lasota2011HLX,Servillat2011,Godet2012,Wu2016}.
Furthermore, some type of non-linear dynamics characteristic for an underlying unstable accretion flow was suggested for a number of other XRBs (Sukova et al. 2015), while the statistical studies of a large sample of sources
support the 'reactivation' scenario in the case of compact radio sources hosting supermassive black holes \citep{Czerny2009}.
On the other hand, many of XRBs and AGN seem to be powered by a stable accretion, despite even large accretion rates. Therefore, some stabilizing mechanisms in the accretion process have been invoked, apart from the viscosity prescription itself. For instance, the propagating fluctuations in the flow
\citep{JaniukMisra2012} may suppress the thermal instability, or at least delay the development of the instability \citep{Ross2017}.
As recently discussed by \citet{Jiang2016}, a possible stabilizing mechanism for the part of an accretion flow might be
the opacity changes connected with the ionization of heavy elements.
Using the shearing-box simulations of MRI-driven fluid in the gravitational potential of supermassive black hole (a particular value of black hole mass, $M= 5 \times 10^{8} M_{\odot}$, was used),
\citet{Jiang2016} have shown that the flow is stable against the thermal instability, if the opacity includes transitions connected with
absorption and scattering on iron ions. This is because the cooling rate, which includes now not only the Thomson scattering (constant) term, but also the absorption and line emission in the Roseland mean opacity, will depend strongly on density and temperature in some specific
regions of the disk.
In fact, as we discuss below in more detail, the dominant term from opacity changes may completely stabilize the flow locally.
The simulations of \citet{Jiang2016} showed that effect but they did not describe the global evolution of the flow, which is the subject of our present work. In our paper we consider the disk stability for the range of black hole masses, characteristic for either XRBs or AGN. We find that the influence of the opacity changes on the global time evolution of the flow is essential, although do not prevent the instability form developing. We show examples of lightcurves produced by our numerical simulation to illustrate this.
The paper is organized as follows: in Sect. \ref{sect:local} we present the analytical condition for local thermal
instability, in Sect. \ref{sect:opal} we present the values of $\kappa$ opacities for the typical accretion disk densities and
temperatures and range of the Iron Opacity Bump, to reveal its influence on global disk behaviour in Sect. \ref{sect:global}.
\section{Local thermal stability in accretion disks}
\label{sect:local}
 The domination of radiation pressure in the accretion disk leads to the thermal instability \citep{PRP1973,LE1974,SS1976,Janiuk2002}.
 At the instability the heating rate $Q_+$ grows faster with temperature than cooling rate $Q_-$.  The appearance of local thermal
instability is given by the condition:
\begin{equation}
 \frac{ d \log Q_-}{d \log T} <  \frac{ d \log Q_+}{d \log T}.
 \label{eq:instability}
\end{equation}
The analysis performed in \citet{viscositypaper}, under the assumption on constant surface density during
thermal timescales leads to following formula on heating rate derivative:
\begin{equation}
 \frac{ d \log Q_+}{d \log T} = 1 + 7 \mu \frac{1 - \beta}{1 +\beta},
  \label{eq:instability2}
 \end{equation}
where $\beta=P_{\rm gas}/P$.
 For the case of this work, we assume $\alpha = 0.02$ and $\mu = 0.56$, being typical for the IMBH and AGN disk case. { We have chosen 
 these values since the dynamics of the outbursts of the disk matches the observed properties of the sources as shown in \citet{viscositypaper}. These values reproduce the correlation between the bolometric luminosity and outburst duration known for the observed sources, especially microquasars and Intermediate Mass Black Holes}.
 The radiative cooling rate depends on disk surface density $\Sigma$ and physical constants Stefan-Boltzmann $\sigma_b$ and speed of light.
 The radiative cooling rate is given by following formula \citep{Janiuk2002,Janiuk2015,viscositypaper}:
 \begin{equation}
  Q_{-} = \frac{4 \sigma_b}{3c \Sigma}  \frac{T^4}{\kappa}.
 \label{eq:qminus}
 \end{equation}
For the radiation pressure dominated disk, combined with Eqs.(\ref{eq:instability}) and (\ref{eq:instability2}), from  Eq. (\ref{eq:qminus}) for $\mu = 0.56$ we get:
\begin{equation}
 \frac{ d \log Q_-}{d \log T} < 4.92.
 \label{eq:qminus2}
\end{equation}
 In our previous papers, only Thomson scattering was assumed, which resulted in the appearance of global radiation pressure instability among all scales of sub-Eddington accretion disks.
The recent results of \citet{Jiang2016} are based on stabilizing influence of the iron opacity components. To confront the results of their short time, local
3D MHD shearing-box simulation, we propose the global model, used previously for the case of Intermediate Mass BH HLX-1 accretion disk \citep{Wu2016,viscositypaper}. According to the lower temperatures in radiation-pressure
dominated areas of the accretion disks, we include in our model also the atomic opacity components. 
Assuming $\kappa$ being function of $\rho$ and $T$, we get the formula for log derivative of
$Q_-$:
\begin{equation}
  \frac{ d \log Q_-}{d \log T} = 4 + \frac{\partial \log \kappa}{\partial \log T} - \frac{4 - 3 \beta}{1 + \beta} \frac{\partial \log \kappa}{\partial \log \rho}.
\label{eq:qminus3}
  \end{equation}
 The Eq. $\ref{eq:qminus3}$ shows possible stabilizing influence of the negative slope of  $\kappa$ dependent on $T$ and destabilizing influence of the positive slope of $\kappa$. Also the dependence on
 $\rho$ has inluence on disk stability.  The value of $\kappa$ itself is not important in the local stability analysis. In case of unstable disk,
 less efficient cooling being an effect of the greater $\kappa$ lowers the temperature of unstable equilibrium solution,  and enlarge the temperature of stable solution, which can modify the duty cycle quantitatively
 but not qualitatively.
 Similarly to \citet{viscositypaper}, we can derive the necessary value for the
 thermal instability:
 \begin{equation}
  \beta < \frac{7 \mu + 3 - \frac{\partial \log \kappa}{\partial \log T} + 4 \frac{\partial \log \kappa}{\partial \log \rho}}{7 \mu - 3 + \frac{\partial \log \kappa}{\partial \log T} - 4 \frac{\partial \log \kappa}{\partial \log \rho}}
 \label{eq:necessarycondition}
 \end{equation}
 We can also define the thermal stability parameter $s$:
 \begin{equation}
 s =  \frac{ d \log Q_+}{d \log T} - \frac{ d \log Q_-}{d \log T} 
  \label{eq:sp}
   \end{equation}
  Using the Eqs. (\ref{eq:instability2}) and (\ref{eq:qminus3}), we can write Eq. (\ref{eq:sp}) as follows:
  \begin{equation}
 s =  - 3 + 7 \mu \frac{1 - \beta}{1 +\beta} - \frac{\partial \log \kappa}{\partial \log T} + \frac{4 - 3 \beta}{1 + \beta} \frac{\partial \log \kappa}{\partial \log \rho}
 \label{eq:sp2}
 \end{equation} 
 The $s$ parameter is connected with the Lyapunov exponent for the system described by the energy  equation in the accretion disk  with stress tensor given by Eq. (1). The value $s > 0$ means locally thermally unstable
 disk, $s \le 0$ - locally thermally stable.  Assuming $\mu = 0.56$ and $\beta << 1$ for $\rho = 10^{-8}$g cm$^{-3}$ and $T = 10^5 $ K we get $s = -11$
 which corresponds to local thermal stability.  Nevertheless, for this density the  parameter $s$ gains positive values for $T > 3 \times 10^5$ K.
Below this temperature, for $T > 1.75 \times 10^5$ K, the disk is locally thermally stable because of the  negative slope of the bump. For temperatures in the range $1.1 - 1.75 \times 10^5$ K, the disk is locally thermally unstable. 
  \section{The variable $\kappa$ - Iron Opacity Bump}
\label{sect:opal}
\begin{figure}
\includegraphics[width=\columnwidth]{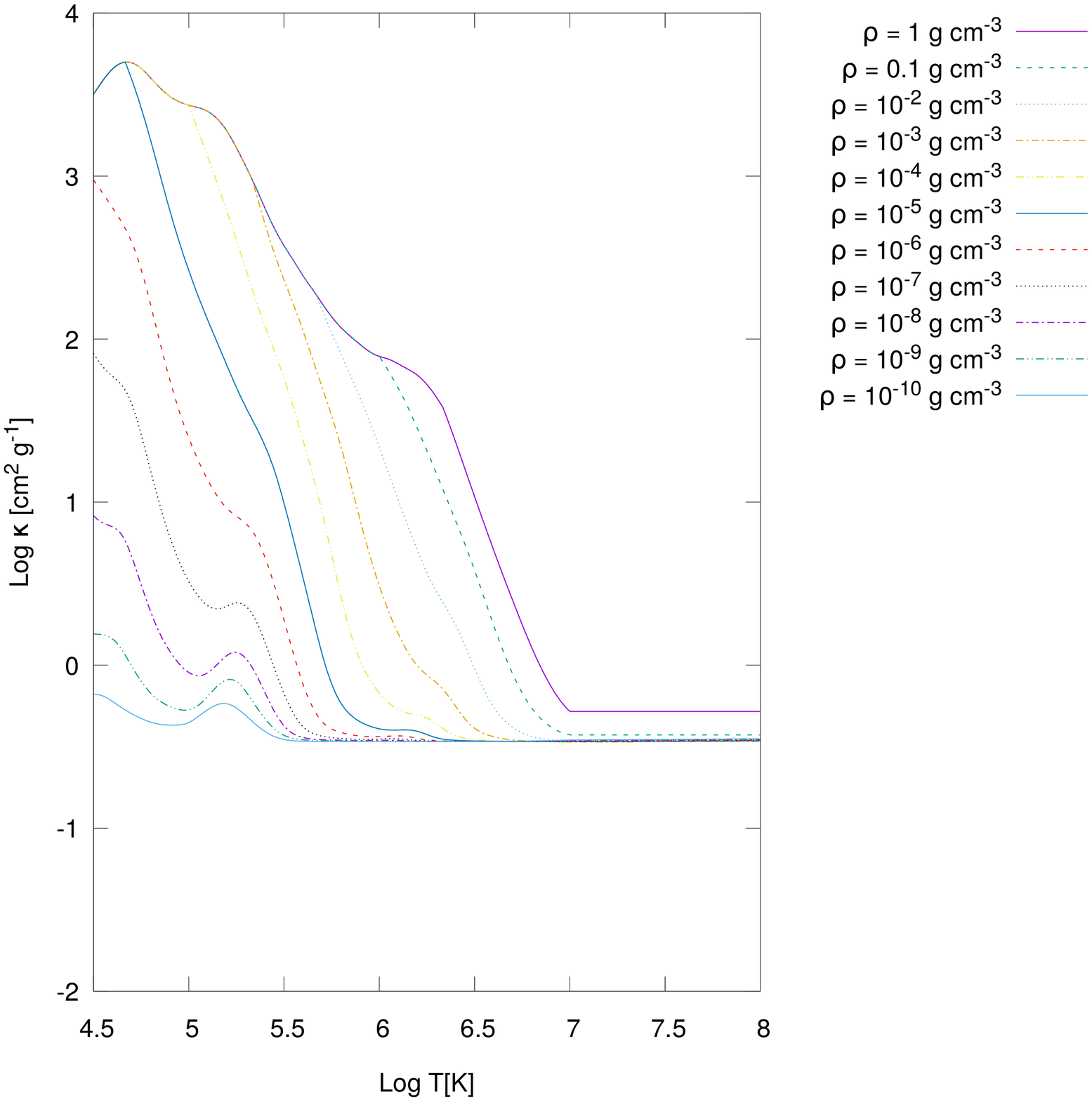}
\caption{The opacity $\kappa$ functions including atomic components. The data are taken from
\citep{Alexander1983,Seaton1994,Rozanska1999}.}
\label{fig:1}
\end{figure}

The opacity $\kappa$ is the local function describing interaction of photons with matter from accretion disks. Under the assumptions of local thermal equilibrium, radiation and gas contribution to the total pressure, and local vertical hydrostatic equilibrium, both the heating and
cooling rates can be described as a function of radius $r$, local density $\rho$ and local temperature $T$. Although the radius $r$, affecting the angular momentum transport is important for the heating rate in the $\alpha$-disk model, it
affects the stability analysis only indirectly, via the parameters of stationary solutions. In Fig. \ref{fig:1} we present the profiles of the total opacity for solar metallicity,
computed as a function of density and temperature \citep{Alexander1983,Seaton1994,Rozanska1999}.
The combined conditions (\ref{eq:instability2}) and (\ref{eq:qminus}) to the opacity values results
in the significant local stabilization for the temperatures of $1 - 4 \times 10^5$ K and densities about $10^{-8}$g cm$^{-3}$
typical for the AGN accretion disks. We fitted the $\kappa$ function with following formulae:
\begin{equation*}
 \kappa = \kappa_{\rm Th} + \kappa_{pl} + \kappa_{\rm bump}
 \label{eq:kappabump}
 \end{equation*},
 \begin{equation*}
 \kappa_{\rm Th} = 0.34 {\rm cm}^2 {\rm g}^{-1},
 \label{eq:kappathom}
 \end{equation*}
\begin{equation}
  \kappa_{pl} =4.6 \times 10^{23} \rho T^{- 3.5},
  \label{eq:kappapow}
 \end{equation}
\begin{eqnarray*}
  \kappa_{\rm bump} = 39.8 \rho^{0.2}(0.8  \exp - (\frac{T - 1.75 \times 10^5 {\rm K} }{8.2 \times 10^4 {\rm K}} )^2 \\+ 6.3 \exp - (\frac{T - 4 \times 10^4 {\rm K} }{3 \times 10^4 {\rm K}} )^2).
  \label{kappabumps}
 \end{eqnarray*}
\begin{figure}
\includegraphics[width=\columnwidth]{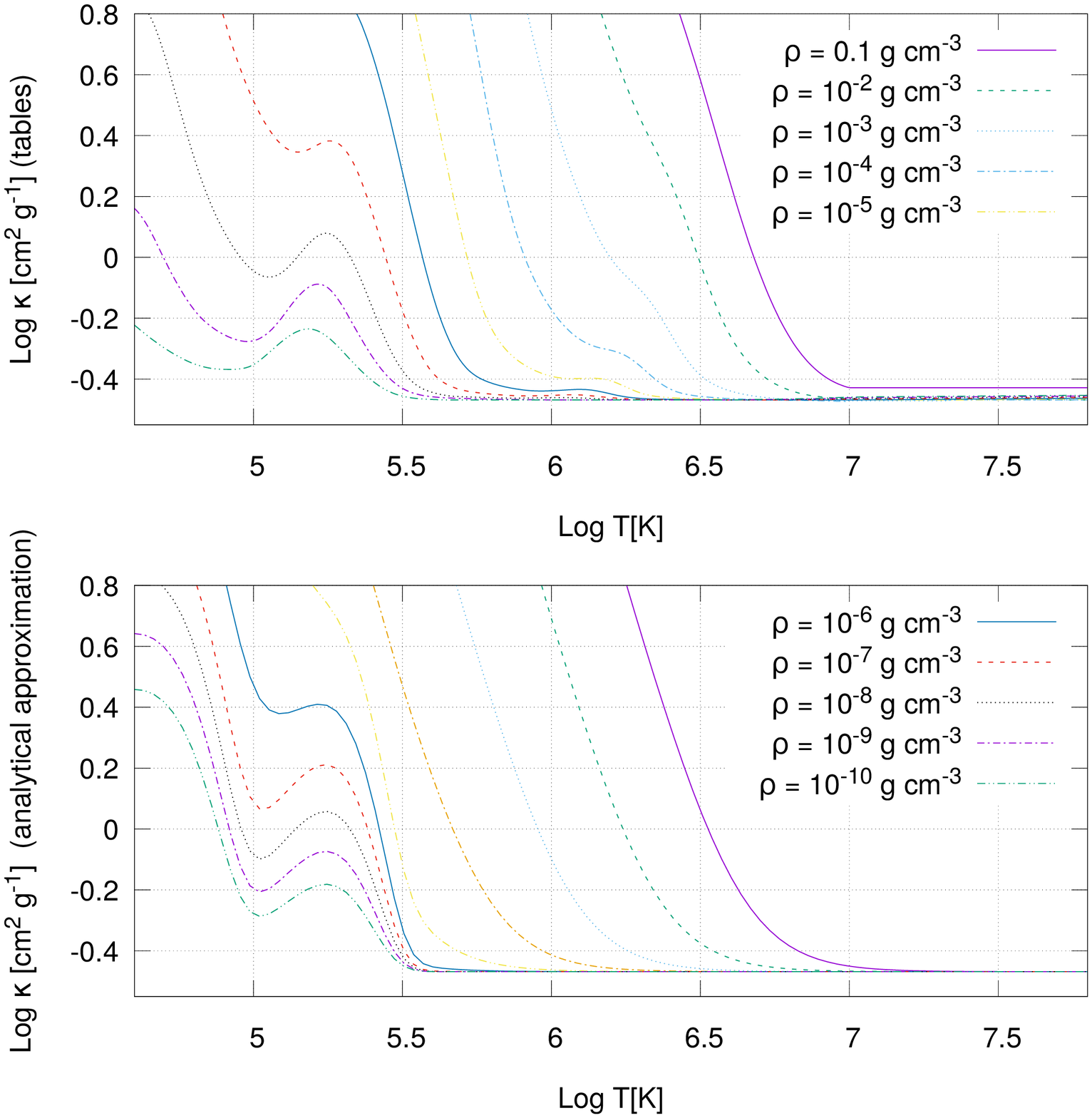}
\caption{Top panel shows results  from the tables \citep{Alexander1983,Seaton1994,Rozanska1999}, and
bottom panel shows the analytical approximation (Eq. \ref{eq:kappapow}).}
\label{fig:2}
\end{figure}
\label{sect:bump}
The negative stabilizing slope of the Iron Opacity Bump is visible in the Figure \ref{fig:1}. In Fig. \ref{fig:2} 
 in the upper panel we show the opacities directly from the tables \citep{Alexander1983,Seaton1994,Rozanska1999}, and in the  lower
panel we present the analytical approximation of opacity function from Eq. (\ref{eq:kappabump}).
The detailed results of the dynamical model are presented in Section \ref{sect:global}.
\begin{figure}
\includegraphics[width=\columnwidth]{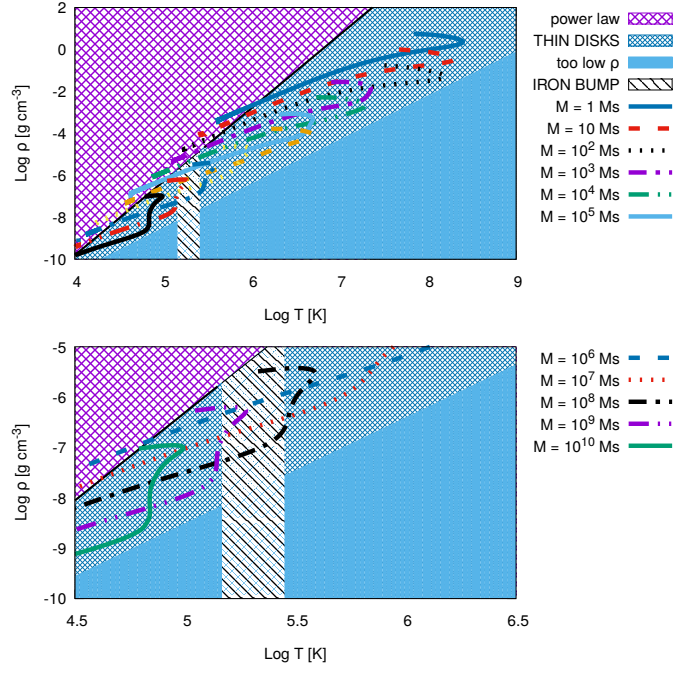}
\caption{Typical values of the $\rho$ and T for Eddingtonian thin disks for central object mass between $1$ and
$10^{10} M_\odot$. We set $\dot m = 0.03$.}
\label{fig:3}
\end{figure}
\section{Global model}
\label{sect:global}
\subsection{Values of $\rho$ and T}
\label{sect:vrht}
In Fig. \ref{fig:3} we present typical values of $\rho$ and T for the wide range of accretion disks, computed via the GLADIS code.
 For the values of $\rho$ and $T$ in the upper left corner, the power law term of $\kappa$ dominates, but matter with this parameter is too dense and too cold for
central areas of sub-Eddington accretion disks. The oblique belt below presents typical values of $\rho$ and $T$ for the accretion disks with different masses. This area, for
$5.1 < \log T < 5.4$, is covered by the Iron Opacity Bump, with stabilizing negative slope (See Sect. \ref{sect:local}).
 According to these results, the stabilizing effect of the negative slope of the Iron Opacity Bump can be visible only for the Active Galactic Nuclei accretion disks with $M \approx 10^8 -
10^9 M_\odot$.
\subsection{Results for the full model with bump}
\label{sect:fullr}
\begin{figure}
\includegraphics[width=\columnwidth]{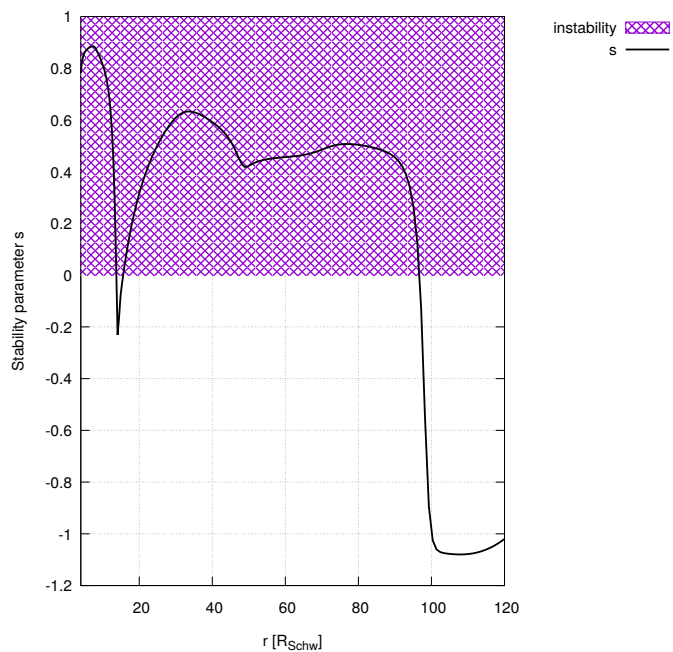}
\caption{The stability parameter, defined in Eq.(\ref{eq:sp}) during the outburst for $M = 5 \times 10^8 M_{\odot}$, $\dot{m} = 0.03$
(same as in \citet{Jiang2016}).}
\label{fig:5}
\end{figure}

We perform the simulations of the global disk behaviour using the time-dependent global code GLADIS \citep{viscositypaper}. The GLADIS code is a time-dependent code, which solves hydrodynamic equations, describing the long-time behaviour
of accretion disk under the assumption of the vertical hydrostatic equilibrium. The code models the time evolution of the flow in
thermal and viscous timescales. Moreover, the code assumes axial symmetry, and Keplerian angular velocity.  In this paper we set $\alpha = 0.02$, $\mu = 0.56$, and $M = 5 \times 10^8 M_\odot$.
The major change in comparison to \citet{viscositypaper} is replacing constant Thomson $\kappa$ with Eq. (\ref{eq:kappabump}).
Similarly to \citet{Jiang2016}, we assumed the Eddington rate $\dot{m} = 0.03$. The results of the time-dependent model are presented in Fig. \ref{fig:5}.
The local shearing-box simulation resulted in the significant stabilization of the disk \citep{Jiang2016}. However, the global model does not confirm
these results. The stabilization of the disk, which appears according to local prediction, is not found in {global models considering a large range of radii}.
Fig. \ref{fig:5} presents the stability parameter profile $s$ defined in Eq. $\ref{eq:sp}$.
For the inner area of the disk, Thomson component of opacity dominates and temperature is too large to expect any form of stabilization. Outer area of the disk characterize the larger value of total opacity
temperature about $1.5 - 2 \times 10^5 K$ and negative values of $s$ (the \textit{bump} temperatures).
The significant gradient of the $s$ is correlated with the gradient of temperature and gradient of $\kappa$ in the
opposite direction.
\begin{figure}
\includegraphics[width=\columnwidth]{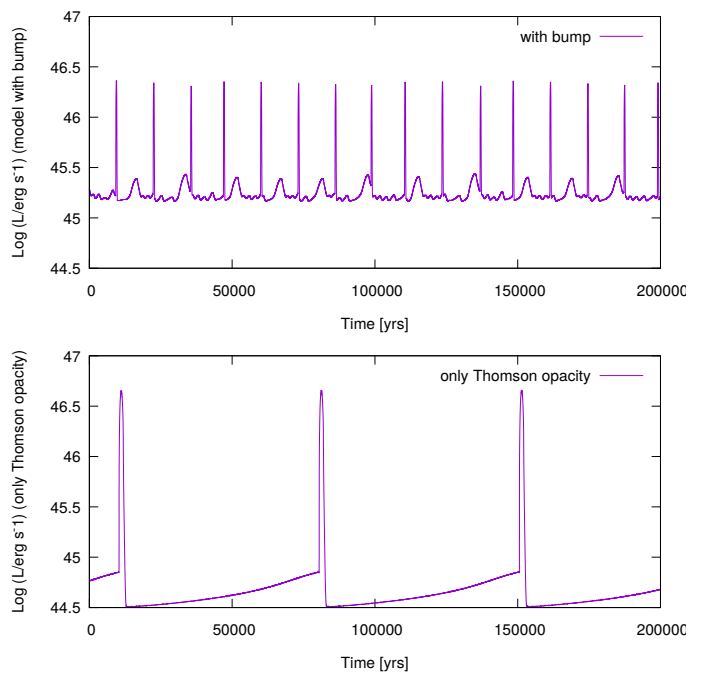}
\caption{Results of the time-dependent model for $M = 5 \times 10^8 M_{\odot}$, $\dot{m} = 0.03$ (same as in \citet{Jiang2016}).
}
\label{fig:6}
\end{figure}
The stability parameter, presented in Fig. \ref{fig:5} can reach values between $-3$ and $-3 + 7 \mu$ ($0.92$ for our choice of $\mu$).
In Fig. \ref{fig:5} the typical profile of the $s$ parameter is presented. As the bump is approximately Gaussian function, centered at $1.75 \times 10^5$ K, with standard deviation $\sigma = 0.82 \times
10^5$ K, it is expected that the strongest effect of the stabilizing slope would be visible for such temperatures. The combined outcome of the stabilizing influence of the negative slope of the bump and destabilizing influence of the positive slope
of the bump for the dynamical model is presented in Fig. $\ref{fig:6}$. In contrast to the results of \citet{Jiang2016}, the iron bump does not stabilize the disk. However, it complicates the lightcurve pattern (many small
short flares preceding main outbursts instead of one simple flare), due to the complexity of the photon absorption process, but the inner regions of the disk remains hot enough to perform the limit-cycle oscillations. In effect,
the bump partially stabilizes the disk - the amplitude $L_{\rm max}/L_{\rm min}$ decreases from 156 for model with constant $\kappa$ from lower panel of \ref{fig:6} to 16.1 for model with bump (upper panel of \ref{fig:6}). The detailed analysis of the lightcurve shape
is presented in Table \ref{tab:tabela}.
\begin{table}
 \begin{tabular}{cccc}
 \hline
 \hline
  model & amplitude A & period P [yrs] & width $\Delta$
 \\ \hline
  $\kappa$ with bump & 16.1& 12635& 0.0023
   \\ \hline
  Thomson $\kappa$ & 156 & 70197 & 0.0044
  \\
 \hline
\end{tabular}
\caption{The table describing the flare parameters for the model from Fig. \ref{fig:6} with $\kappa$ described by formula
$(\ref{eq:kappabump})$ ( second row) and only Thomson $\kappa$ (third row). Those parameters were described more precisely in \citet{viscositypaper}.
}
\label{tab:tabela}
\end{table}
 Since the timescales presented in Fig. \ref{fig:6} are much longer than the duration of observation (lasting up to
several decades), it is impossible to find such a lightcurve using the direct method. For such a black hole mass, 
during the phase of fastest growth of the luminosity, the luminosity change can reach 1 per cent per year. The shape of the lighturve can be reflected in the 
Eddington rate statistic - similar objects, being in the different phase of the limit-cycle presented in Fig. \ref{fig:5} can emit the radiation with different luminosity and spectra.
However, in case of much smaller black hole masses new timescales are perhaps accessible to observations. For 
example, digitalization of Harvard plates \citep{Grindlay2012} \footnote{http://dasch.rc.fas.harvard.edu/} will bring lightcurves on the order of a hundred years and perhaps the
outbursts of AGN disks can be discovered.
\section{Conclusions}
In our previous paper \citep{viscositypaper} we computed large grids of models confirming the universality of radiation pressure instability across the BH mass-scale. In this paper we
changed the opacity prescription to examine the heavy atoms influence on the accretion disk instability. Comparison between
two models presented in Table \ref{tab:tabela} leads to the conclusions that heavy atoms stabilize the disk partially, but
do not imply that the variability vanishes. This stabilizing effect manifests itself rather in a significant period and amplitude decrease, without a relative broadening of the outbursts with respect to their separation .
Additionally, some mild \textit{precursors}, being an outcome of a non-monotonic profile of the $s$ parameter distribution, are also visible.
That partial stabilization, being an important effect for the Active Galactic Nuclei, has only weak influence on the radiation pressure
among all BH mass-scale in accretion disk under the assumption of solar metallicity. In case of sources with different metallicity,
this effect can change its extent.  Finally, we conclude that the radiation pressure driven limit cycle oscillations, suffering some disturbances from the \textit{Iron Opacity Bump} in case of the AGN disks are also expected,
at least for moderately large supermassive black holes ($M = 5 \times 10^8 M_\odot$).
\section*{Acknowledgments}
This work was supported in part by the grants DEC-2012/05/E/ST9/03914 and 2015/18/M/ST9/00541 from the Polish National Science Center.


\begin{thebibliography}{}
\bibitem[Abramowicz et al. (1988)]{Abramowicz1988} {{Abramowicz}, M.~A., {Czerny}, B., {Lasota}, J.~P.,{Szuszkiewicz}, E.} E.~J., et al.\ 1988, \apj, 332, 646
\bibitem[Alexander et al.(1983)]{Alexander1983} {Alexander}, D.~R., {Rypma}, R.~L., {Johnson}, H.~R. \ 1983, \apj, 272, 773
\bibitem[Altamirano et al. (2011)]{Altamirano2011} {Altamirano}, D., {Belloni}, T., {Linares}, M., {van der Klis}, M. et al. \ 2011, \apjl, 742, L17
\bibitem[Belloni et al. (2000)]{Belloni2000} {{Belloni}, T., {Klein-Wolt}, M., {M{\'e}ndez}, M., 
	{van der Klis}, M., {van Paradijs}, J.} \ 2000, \aap, 335, 271
\bibitem[Blaes (2014)]{Blaes2014}	{Blaes}, O.\ 2014, \ssr, 183, 221
\bibitem[Czerny et al. (2009)]{Czerny2009} {{Czerny}, B., {Siemiginowska}, A., {Janiuk}, A., {Nikiel-Wroczy{\'n}ski}, B., 
	{Stawarz}, {\L}.} \ 2009, \apj, 698, 840
\bibitem[Farrell et al. (2009)]{Farrell2009} {{Farrell}, S.~A., {Webb}, N.~A., {Barret}, D., {Godet}, O., 
	{Rodrigues}, J.~M.} \ 2009, \nat, 460, 73
\bibitem[Godet et al. (2012)]{Godet2012} {Farrell}, S.~A., {Webb}, N.~A., {Barret}, D., {Godet}, O., 	{Rodrigues}, J.~M. \ 2012, \apj, 752, 34
\bibitem[Grindlay et al. (2012)]{Grindlay2012} Grindlay, J., Tang, S., Los, E. and Servillat, M.
	 \ 2012, New Horizons in Time Domain Astronomy, Proceedings of the International Astronomical Union, 285, 29
\bibitem[Grz{\c e}dzielski et al. (2017)]{viscositypaper} {Grz{\c e}dzielski}, M., {Janiuk}, A., {Czerny}, B., 
	{Wu}, Q. \ 2017, in press, ArXiv e-prints, 1609.09322
\bibitem[Janiuk, Czerny \& Siemiginowska (2002)]{Janiuk2002} Janiuk, A., Czerny, B., Siemiginowska, A. \ 2002, \apj, 576, 908
\bibitem[Janiuk et al. (2015)]{Janiuk2015} {Janiuk}, A., {Grz{\c e}dzielski}, M., {Capitanio}, F.,{Bianchi}, S. \ 2015, \aap, 574
\bibitem[Janiuk \& Misra (2012)]{JaniukMisra2012} {Janiuk}, A., {Misra}, R., \aap, 540, 114
\bibitem[Jiang, Davis \& Stone (2016)]{Jiang2016}  {Jiang}, Y.-F., {Davis}, S.~W., {Stone}, J.~M. \ 2016, \apj, 827, 10
\bibitem[Lasota et al. (2011)]{Lasota2011HLX} {Lasota}, J.-P., {Alexander}, T., {Dubus}, G., {Barret} et al. \ 2011, \apj, 735, 89
\bibitem[Lightman \& Eardley (1974)]{LE1974} {Lightman}, A.~P., {Eardley}, D.~M.\ 1974, \apjl, 187, 1
\bibitem[Pringle, Rees \& Pacholczyk (1973)]{PRP1973} {Pringle}, J.~E., {Rees}, M.~J., {Pacholczyk}, A.~G. \ 1973, \aap, 29, 179
\bibitem[Ross, Latter \& Tehranchi (2017)]{Ross2017} {Ross}, J., {Latter}, H., {Tehranchi}, M. \ 2017, arXiV e-prints, 1703.00211 
\bibitem[Rozanska et al. (1999)]{Rozanska1999} {R{\'o}{\.z}a{\'n}ska}, A., {Czerny}, B., {{\.Z}ycki}, P.~T.,{Pojma{\'n}ski}, G. \ 1999, \mnras, 305, 491
\bibitem[Seaton et al. (1994)]{Seaton1994} {Seaton}, M.~J., {Yan}, Y., {Mihalas}, D., {Pradhan}, A.~K. \ 1994, \mnras, 266, 805
\bibitem[Servillat et al. (2011)]{Servillat2011} {Servillat}, M., {Farrell}, S.~A., {Lin}, D., {Godet}, O., {Barret}, D., {Webb}, N.~A. \ 2011, \apj, 743, 6
\bibitem[Shakura \& Sunyaev (1973)]{SS1973} {Shakura}, N.~I., {Sunyaev}, R.~A. \ 1973, \aap, 24, 337
\bibitem[Shakura \& Sunyaev (1976)]{SS1976} {Shakura}, N.~I., {Sunyaev}, R.~A \ 1976, \mnras, 175, 613
\bibitem[Szuszkiewicz (1990)]{Szuszkiewicz1990} {Szuszkiewicz}, E. \ 1990, \mnras, 244, 377
\bibitem[Wu et. al (2016)]{Wu2016} {Wu}, Q., {Czerny}, B., {Grz{\c e}dzielski}, M., {Janiuk} et al.\ 2016, \apj, 833, 79
\end{thebibliography}
\end{document}